\documentclass[apj]{emulateapj}

\newcommand{\kms}{{\rm km~s\ensuremath{^{-1}}}}
\newcommand{\msigma}{$M_\bullet - \sigma_\ast$}
\newcommand{\msigmau}{$(M_\bullet - \sigma_\ast)_u$}
\newcommand{\bh}{$M_\bullet$}
\newcommand{\sig}{$ \sigma_\ast$}

\usepackage{natbib}
\usepackage{multirow}
\usepackage{appendix}
\usepackage{longtable}

\slugcomment{ApJL Accepted}

\shorttitle{Spheres of Infuence and the  \msigma\ relation}
\shortauthors{Batcheldor}

\begin{document}


\title{The \msigma\ Relation Derived from Sphere of Influence Arguments}


\author{D. Batcheldor}
\affil{Department of Physics and Space Sciences, Florida Institute of Technology, \\150 West University Blvd, Melbourne, FL, 32901, USA}
\email{dbatcheldor@fit.edu}



\begin{abstract}
The observed relation between supermassive black hole (SMBH) mass (\bh) and bulge stellar velocity dispersion (\sig) 
is described by $\log{M_\bullet} = \alpha + \beta\log{(\sigma_\ast/200\kms)}$. As this relation has important implications 
for models of galaxy and SMBH formation and evolution, there continues to be great interest in adding to the \bh\ catalog.
The ``sphere of influence'' ($r_i$) argument uses spatial resolution to exclude some \bh\ estimates and pre-select additional 
galaxies for further SMBH studies. This {\it Letter} quantifies the effects of applying the $r_i$ argument to a population 
of galaxies and SMBHs that do not follow the \msigma\ relation. All galaxies with known values of \sig, closer than 100~Mpc, 
are given a random \bh\ and selected when $r_i$ is spatially resolved. These random SMBHs produce a \msigma\ relation of 
$\alpha=8.3\pm0.2,\beta=4.0\pm0.3$, consistent with observed values. Consequently, future proposed \bh\ estimates should not 
be justified solely on the basis of resolving $r_i$. This {\it Letter} shows the observed \msigma\ relation may simply be a 
result of available spatial resolution. However, it also implies the observed \msigma\ relation defines an upper limit. 
This potentially provides valuable new insight into the processes of galaxy and SMBH formation and evolution.
\end{abstract}




\section{Introduction}

There has been a growing database of direct supermassive black hole (SMBH) mass (\bh) estimates from the centers of nearby galactic bulges 
\citep[e.g.,][]{2008PASA...25..167G}. While the limits of our current abilities to significantly expand this database may have been reached 
\citep{2009PASP..121.1245B}, the last decade has seen a wealth of \bh\ estimates that have increased the SMBH catalog from 13 or 26 
\citep{2000ApJ...539L...9F,2000ApJ...539L..13G} to $\sim$70 \citep{2008PASA...25..167G,2008MNRAS.386.2242H,2009ApJ...698..198G}. An 
intense interest in populating the SMBH database was sparked by observed correlations between \bh\ and fundamental properties of their 
host bulges \citep[e.g.,][]{1995ARA&A..33..581K,1998AJ....115.2285M,2000ApJ...539L...9F,2000ApJ...539L..13G,2001ApJ...563L..11G,
2002ApJ...578...90F,2003ApJ...589L..21M,2003MNRAS.341L..44B,2004ApJ...604L..89H,2005ApJ...631..785P}. These \bh\ scaling relations 
have generated numerous theoretical investigations \citep[e.g.,][]{2001ApJ...552L..13C,2003ApJ...591..125A,2005MNRAS.364..407C,
2006ApJ...641...90R} and have possibly added valuable limits to evolutionary models \citep[e.g.,][]{2004ApJ...613..109H,
2005ApJ...634..910W,2007ApJ...667..117T,2008arXiv0808.1349C}. 

\begin{figure}
\plotone{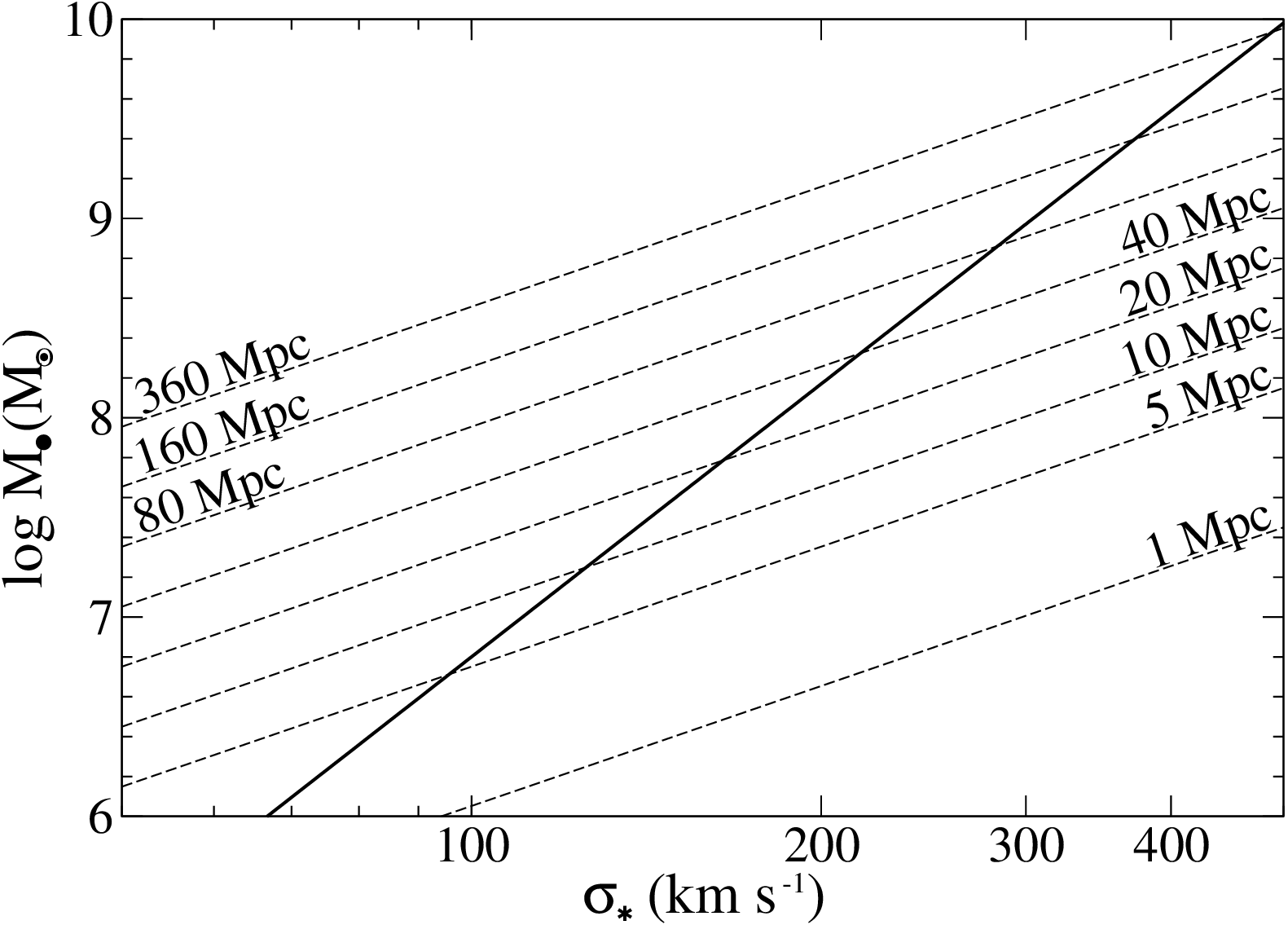}
\caption{{\it Observable} \msigma\ plane at discrete distances based on $\Re=0\farcs1$ and $\alpha=8.2,\beta=4.6$ (solid line). In areas 
above the dashed distance lines, $r_i$ will be resolved if the galaxy with the given \sig\ hosts the given SMBH. 
}
\label{fig:1}
\end{figure}

The degree to which a SMBH's sphere of influence, $r_i$, is resolved has been used as a quality measure for \bh\ estimates 
\citep{2002ApJ...578...90F,2003ApJ...589L..21M,2004ApJ...602...66V}. The only method available to a priori determine if a 
\bh\ estimate can be made is to assume $r_i = GM_\bullet/\sigma_\ast^2$ \citep{1972ApJ...178..371P}. However, to calculate 
$r_i$ in galaxies with known \sig, \bh\ is estimated using the \msigma\ relation given by $\log{M_\bullet} = \alpha + 
\beta\log{(\sigma_\ast/200\kms)}$, where $\alpha=8.1-8.2$ and $\beta=3.7-4.9$. The observed scatter, $\epsilon$, is 0.4~dex 
\citep{2006ApJ...637...96N,2009ApJ...698..812G}. Following this, Figure~\ref{fig:1} demonstrates where $r_i$ will be resolved, 
given a spatial resolution of $\Re=0\farcs1$. A value of $\Re=0\farcs1$ is used here as that is the typical FWHM of the {\it HST} 
PSF. To date {\it HST} has been responsible for most \bh\ estimates. 

\citet[][FF05]{2005SSRv..116..523F} discussed the limited abilities of {\it HST} to resolve $r_i$, and \citet[][G09]{2009ApJ...698..198G} 
found the \msigma\ relation to be biased when applying the $r_i$ argument. The influence of $r_i$ cuts on the \msigma\ relation is continued 
in this {\it Letter} with two significant advances. First, a sample of all galaxies with a known \sig\ ($<$100 Mpc) is used. It is important 
to only use these galaxies as there are no \sig=400\kms galaxies at 1~Mpc, for example. Second, random \bh\ estimates are applied to each 
galaxy, i.e., no galaxy is assumed to intrinsically fall on the \msigma\ relation. This ensures no pre-selection of galaxies that 
lie on the \msigma\ relation. Throughout, a distinction is made between the {\it observed} \msigma\ relation (published values) and the 
{\it observable} \msigma\ relation (the relation that can be fitted using the data simulated here). 

\section{Methods and Results}\label{mands}

In short, these simulations take a galaxy with a known \sig\ and distance, assign a \bh, then calculate $r_i$. The $r_i$ argument is then 
applied; if $r_i$ is unresolved the galaxy is removed from the sample (the low mass cut). If the assigned \bh\ generates a galaxy that lies 
above the {\it observed} \msigma\ relation, it is also removed from the sample (the high mass cut). The {\it observable} \msigma\ relation 
is then fit to the remaining galaxies using a Levenberg-Marquardt least-squares add-on to IDL. 

An SQL search of the Hyperleda\footnote{http://leda.univ-lyon1.fr/} catalog was performed \citep{2003A&A...412...45P}. All galaxies with a 
kinematical distance modulus less than 35.0 (100~Mpc) were found. Of the 49740 galaxies returned, 2518 
have published values of \sig. The incompleteness of this sample, due to difficulties in measuring \sig\ in faint galaxies, is noted. Values 
of $r_i$ (in arcsecs) were then calculated for each galaxy by assigning a \bh.

In the first cases, the low mass cut was made by applying $\Re=0\farcs1$ to represent the {\it HST} PSF. However, several values of $\Re$ are 
applied later. The high mass cuts were then made by considering the {\it observed} distribution of galaxies in the \msigma\ plane; there is a clear 
absence of over-massive SMBHs, i.e., there are no \sig=50\kms\ dwarf spheroids in the Local Group with \bh$\sim10^9M_\odot$. Such a nearby over-massive 
SMBH population would have been detected, and therefore it is assumed that these objects do not exist. Galaxies selected for the high mass cuts 
were determined by assigning a scatter ($\epsilon=0.4$~dex) and upper limits to the \msigma\ relation, \msigmau. Galaxies with 
$r_i > r_{\rm max}(M_{\rm max})$ were removed from the sample, where $M_{\rm max} = [\epsilon + 10^{\alpha_u}(\sigma/200\kms)^{\beta_u}]M_\odot$. 

\begin{figure}
\plotone{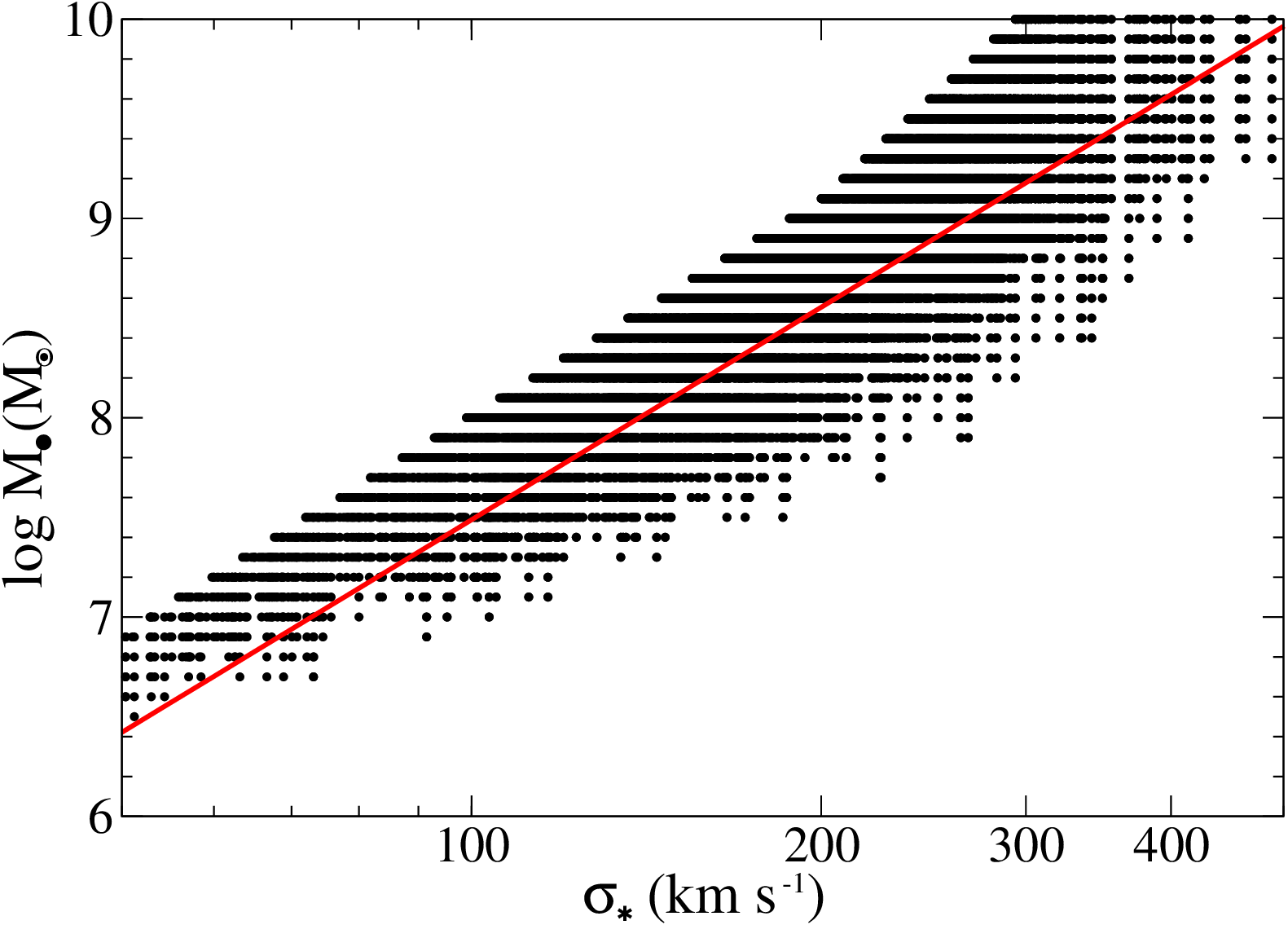}
\caption{Simulated \msigma\ data points (black circles) based on a $\Re=0\farcs1$ criteria. This demonstrates the {\it observable} region of 
the \msigma\ plane with the assumption that there is no \msigma\ relation. The fitted \msigma\ relation (red line) is $\alpha = 8.4, \beta = 3.5$. 
}
\label{fig:2}
\end{figure}

Each galaxy was first assigned 90 separate values of \bh\ using a step size of $10^{0.1}$ from $10^{1.0}M_\odot$ to $10^{10.0}M_\odot$. In this 
case, the high mass cut was made by applying the $\alpha_u=8.1, \beta_u=4.2$ relation of G09 and the $\alpha_u = 8.2, \beta_u = 4.9$ relation 
of FF05. Figure~\ref{fig:2} shows the distribution of {\it observable} galaxies in the \msigma\ plane assuming every galaxy within 100~Mpc can 
host any \bh\, and that $\Re=0\farcs1$. The fit to this sample (red line) is $\alpha = 8.4, \beta = 3.5$.

\begin{figure}
\plotone{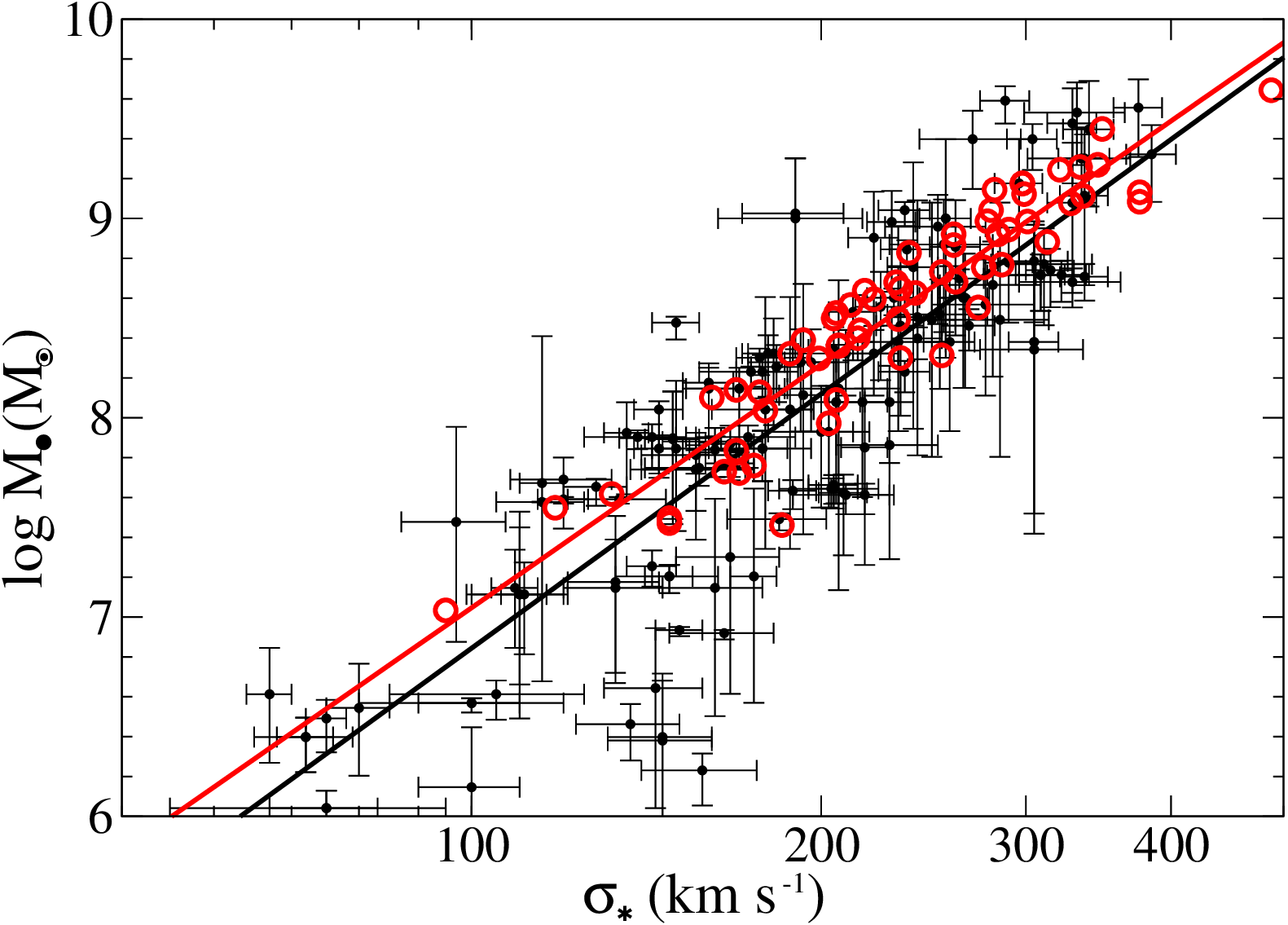}
\caption{Distribution of {\it observable} galaxies from a random distribution of \bh\ (red circles) using the \msigmau\ relation of G09. 
All present estimates of \bh\ and \sig\ are shown as filled black circles, with uncertainties. The fit to the {\it observable} galaxies (red circles and line) 
is $\alpha=8.3,\beta=4.1$
}
\label{fig:3}
\end{figure}

However, galaxies host single (or binary) SMBHs, therefore it is shrewd to assign each galaxy a single (uniformly sampled) random value for \bh\ in the 
range $10^1 - 10^{10}M_\odot$. The distribution of {\it observable} galaxies in the \msigma\ plane using a random sample of \bh\ is shown in 
Figure~\ref{fig:3} using the red open circles. In this case, the low cut was made using $\Re=0\farcs1$, and the high mass cut was made using a 
\msigmau\ relation of $\alpha_u=8.1, \beta_u=4.2, \epsilon=0.4$ (G09). The \msigma\ relation fit to these {\it observable} galaxies (red line) is 
$\alpha_u=8.3, \beta_u=4.1, \epsilon=0.2$~dex. An {\it observable} \msigma\ relation of $\alpha_u=8.3, \beta_u=4.6, \epsilon=0.4$~dex is found 
using the high mass \msigmau\ relation of FF05. Figure~\ref{fig:3} also shows all {\it observed} galaxies based on the combined catalogs of 
\cite{2008PASA...25..167G}, \cite{2008MNRAS.386.2242H} and G09. No distinction is made between ``good'' and ``bad'' \bh\ estimates, or differences in 
quoted \sig; all estimates are plotted (143 total). 

\begin{deluxetable}{lccc}
\tablecaption{Effects of the $r_i$ Cutoff From Averaged Random \bh\ Distributions \label{tab:1}}
\tablewidth{0pt}
\tablehead{
\colhead{$\Re$}&
\colhead{Observed $\alpha$}&
\colhead{Observed $\beta$}&
\colhead{Observed $\epsilon$} \\
\colhead{}&
\colhead{(Zero-point)}&
\colhead{(Slope)}&
\colhead{(Scatter)} \\
}
\startdata

\multicolumn{4}{c}{Variable Upper Cutoffs}\\ \hline
0\farcs05 & $8.25\pm0.17$ & $3.80\pm0.33$ & 0.41 dex \\
0\farcs10 & $8.25\pm0.21$ & $3.90\pm0.34$ & 0.27 dex \\
0\farcs15 & $8.25\pm0.23$ & $3.93\pm0.34$ & 0.22 dex \\
0\farcs20 & $8.25\pm0.27$ & $4.01\pm0.38$ & 0.18 dex \\ \hline \\
\multicolumn{4}{c}{Fixed Upper Cutoff}\\ \hline
0\farcs05 & $8.44\pm0.03$ & $3.93\pm0.23$ & 1.82 dex \\
0\farcs10 & $8.57\pm0.02$ & $4.02\pm0.21$ & 1.22 dex \\
0\farcs15 & $8.64\pm0.02$ & $4.03\pm0.25$ & 0.97 dex \\
0\farcs20 & $8.68\pm0.03$ & $4.07\pm0.32$ & 0.85 dex \\
\enddata
\tablecomments{Average results from applying variable and fixed \msigma\ upper limits to 50 random \bh\ samples. A fixed upper limit of 
$\alpha_u=8.7$ and $\beta_u=5.0$ was used based on the upper limit fit in Figure~\ref{fig:4}. If there was no intrinsic \msigma\ relation, 
these are the values of the \msigma\ relation that would be observed using the different spatial resolutions listed.  
}
\end{deluxetable}

The similarity between the \msigma\ distribution of {\it observable} random mass SMBHs and {\it observed} SMBHs masses is striking. However, as 
this result derives from a single random sampling of \bh, it represents a single possible {\it observable} \msigma\ relation. If galaxies 
intrinsically have random \bh\ values, then the range of {\it observable} SMBHs could have a distribution given by Figure~\ref{fig:2}. 
Therefore, 50 separate random \bh\ samplings were then made for each galaxy, and in each case a fit to the {\it observable} \msigma\ relation was 
performed. In addition, as the previously used value of $\Re=0\farcs1$ only applies to {\it HST} observations, the analysis was repeated for a range 
of $r_i$ lower cut offs. Finally, as the high mass cuts may not actually be defined by the G09 and FF05 fits, values of $\alpha_u$ from 7.8 to 8.7 
(in 0.1 steps) and $\beta_u$ from 3.6 to 5.4 (in 0.2 steps) were used to create an addition 100 individual \msigmau\ cutoffs. The mean values of 
$\alpha,\beta$ and $\epsilon$, derived by applying these different $\Re$ criteria, and from using these variable upper limits, are presented in 
Table~\ref{tab:1}. 

\begin{figure}
\plotone{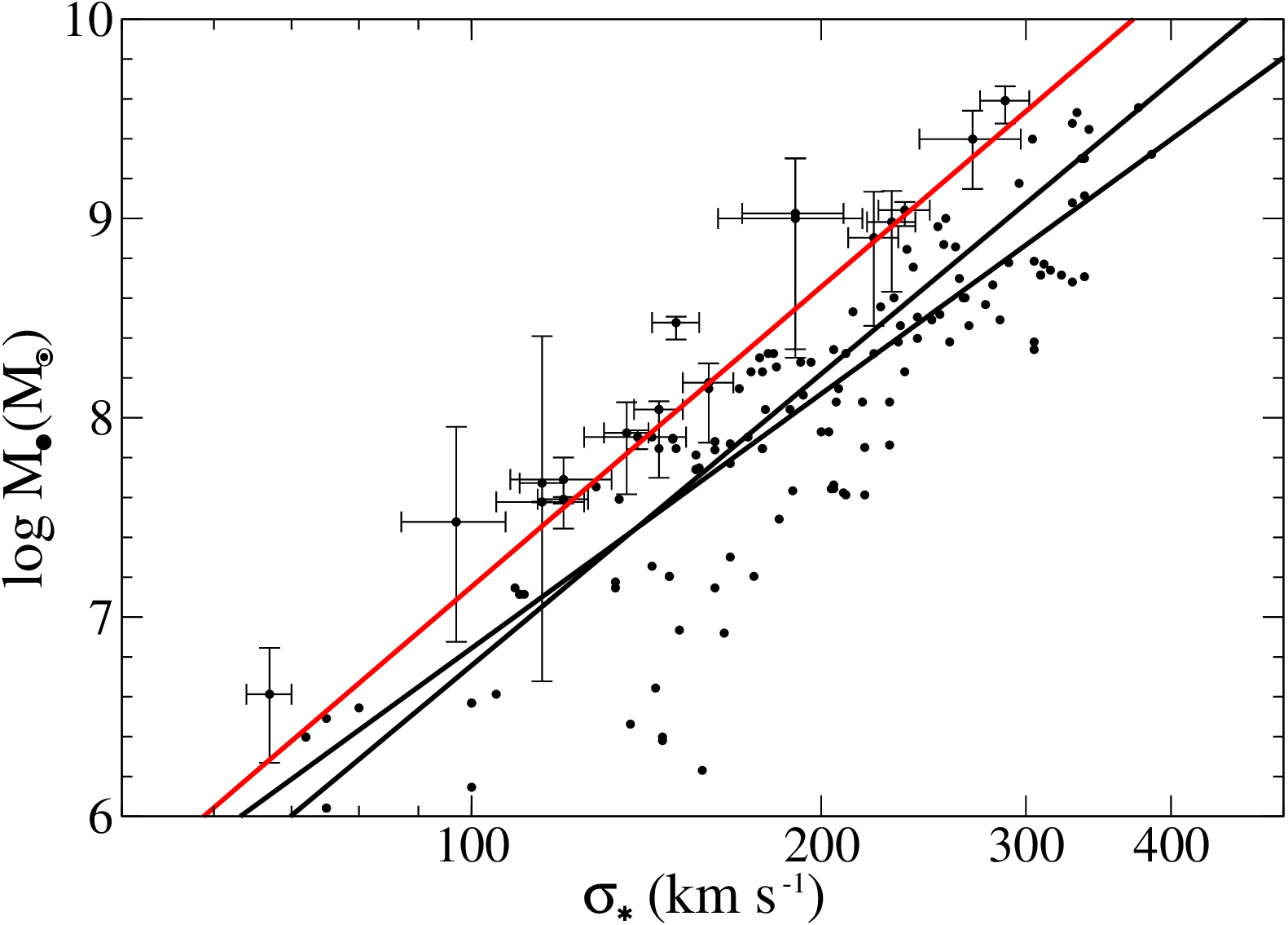}
\caption{Fitting the observed \msigmau\ relation. Filled black circles are the \msigma\ points, those with error bars are the points used for the upper fit 
(shown in red). The red upper limit fit is $\alpha=8.7, \beta=5.0$. The two solid black lines are the {\it observed} G09 and FF05 fits. 
}
\label{fig:4}
\end{figure}

Table~\ref{tab:1} has several notable features. First, as $\Re$ rises $\epsilon$ falls. This is expected because the range of {\it observable} SMBHs 
decreases with larger values of $\Re$. Second, the slope ($\beta$) remains consistent. Finally, in the case of the variable upper limits, the zero-point 
($\alpha$) remains constant and is consistent with the mid-value of the \msigmau\ limits imposed on the sample. To test whether these values of $\alpha$ 
are a consequence of the high mass cut conditions, the analysis was repeated using $\alpha_u$ limits of 7.0 and 8.8 (in 0.2 steps). Applying a $\Re=0\farcs1$ 
low mass cut, an {\it observable} relation with $\alpha=8.0\pm0.5$ was found. Again, the values of $\alpha$ are consistent with the mid-value of the imposed 
limits. However, an estimate of the {\it observed} \msigmau\ relation can be made. The {\it observed} SMBH sample in Figure~\ref{fig:3} was used to define 
an upper limit by fitting all {\it observed} SMBHs that fall above the FF05 and G09 relations, and above $\epsilon=0.4$. The fit to these over-massive SMBHs 
produce an upper limit of $\alpha_u=8.7, \beta_u=5.0$ (Fig.~\ref{fig:4}). This \msigmau\ fit was then applied to 50 random values of \bh\ in each galaxy, and 
the {\it observable} \msigma\ relations were then fitted in each case. The average of these fits are also presented in Table~\ref{tab:1}. In this case, the 
fitted values of $\alpha$ and $\beta$ are consistent with the variable upper limits, but the values of $\epsilon$ are significantly larger. This is expected, 
as these fits were applied to data that had the maximum range in the \msigma\ plane due to the high {\it observed} values of $\alpha_u$ and $\beta_u$.

\section{Discussions}

In summary, as a result of applying the $r_i$ argument, a \msigma\ relation consistent with {\it observed} values ($\alpha\approx8, \beta\approx4, 
\epsilon\approx0.3$~dex) can be fitted to a sample of galaxies that contain random mass SMBHs, and as a consequence do not follow a \msigma\ relation. 
The $r_i$ argument removes low mass SMBHs where $r_i$ would not be resolved, and high mass SMBHs where $r_i$ should have been resolved if such a population 
were present. Therefore, for scaling relations to be of value in constraining galaxy evolution models, \bh\ estimates must not be solely proposed on the 
basis of critically resolving values of $r_i$ derived from \sig. 

It is unlikely that an intrinsic \msigma\ relation has been {\it observed} as a result of critically sampling $r_i$ in the \msigma\ plane, and it will be 
some time until we are able to sample a significant area rightward of the {\it observed} \msigma\ relation \citep[e.g., FF05;][]{2009PASP..121.1245B}. 
Therefore, it is clear that caution must be used in the application of the {\it observed} \msigma\ relation until this issue can be resolved. 

The possibility of there being no tight intrinsic \msigma\ relation, and that the distribution of {\it observed} galaxies in the \msigma\ plane
determine an upper limit only, is now discussed. As the same \bh\ estimates are used to define the other \bh\ scaling relations, this also implies that 
all {\it observed} scaling relations are upper limits. There are a number of questions arising from this possibility, beginning with the most fundamental. 
Why do {\it observed} SMBHs fall on the \msigma\ relation? There are several SMBHs with such high quality \bh\ estimates (Sgr A*, NGC~4258, M87), that there 
clearly is a relation between \bh\ and \sig\ in some galaxies. However, the \msigma\ plane has been increasingly populated with less certain \bh\ estimates 
that have potentially been included on the assumption that $r_i$ is spatially resolved, i.e., that they follow a previously estimated \msigma\ relation 
defined by higher quality \bh\ estimates. In these cases, as the simulations presented here show, an {\it observed} \msigma\ relation will arise simply 
as a result of the $r_i$ selection effect, even if \bh\ is randomly distributed within galaxies. 

Does a population of under-massive SMBHs exist, and have they been detected? If the \msigma\ relation is an upper limit, then there should be galaxies 
that host low mass SMBHs, as suggested by simulations \citep{2005MNRAS.363.1376V,2007ApJ...663L...5V}. Indeed, both the simulations presented 
here, and the observations plotted in Figure~\ref{fig:3}, show that under-massive SMBHs can be, and have been, detected. Some notable cases are that 
of NGC~4435 \citep{2006MNRAS.366.1050C}, a sample of barred galaxies \citep{2008ApJ...680..143G} and narrow-line Seyfert 1 galaxies 
\citep{2005ApJ...633..688M}. In fact, the evidence to suggest the presence of under-massive SMBHs is not matched by any evidence for a significant 
population of over-massive SMBHs leftward of the {\it observed} \msigma\ relation.

It is important to note some intricacies with the two dominant techniques for measuring \bh\ (stellar and gas dynamics). In the case of stellar dynamics, 
there are systematics to the models that may allow a large range of \bh\ in a given bulge \citep{2004ApJ...602...66V}. In the case of gas dynamics, it is 
unclear what the inclination of the nuclear gas disk may be \citep[e.g.,][]{2003ApJ...586..868M}. Both methods allow the potential for many of the current 
\bh\ estimates to in fact be upper limits, i.e., the true \msigma\ plane may have a large distribution of under-massive SMBHs. Therefore, under-massive 
SMBHs may have been observed, their mass over-estimated, and their impact over-looked. 

All SMBH models allow stringent upper limits to be placed on \bh\ \citep[e.g.,][]{2002ApJ...567..237S,2009ApJ...692..856B}. However, these \bh\ upper limits will also 
be dependent on the available spatial resolution. Any kinematical data will have two velocity points spatially separated on a scale of $\Re$. An upper limit to \bh\ is 
estimated by including an increasing dark mass until the derived model becomes inconsistent with the data, i.e., a higher upper limit to \bh will be estimated using a lower 
$\Re$. There are still important constraints that can be added to the \msigma\ plane from estimates of \bh\ upper limits, however, as upper limits that fall below the 
{\it observed} \msigma\ relation provide the same evidence for under-massive SMBHs as would a tightly constrained low mass \bh. At present, most \bh\ upper 
limits are based on data derived from gas dynamics. Consequently, these limits generally fall above the {\it observed} \msigma\ relation due to unknown amounts 
of line broadening from non-gravitational processes, and uncertainties in the inclination if the nuclear gas disk. 

If the \msigma\ relation represents an upper limit in the \msigma\ plane, then what is this limit? In \S~\ref{mands} an upper limit of $\alpha_u=8.7, \beta_u=5.0$ was 
found based on the distribution of {\it observed} SMBHs leftward of the {\it observed} \msigma\ relation. This is likely a good approximation to the \msigmau\ limit 
as there are no reports of a steeper relation. However, this estimate does not include the potential over-massive SMBHs from the upper limits calculated by 
\cite{2009ApJ...692..856B}. Including these limits to the upper limit sample gives values of $\alpha_u=8.8, \beta_u=3.9$ to \msigmau. As expected, due to the 
addition of SMBH limits at lower \bh, this \msigmau\ limit is more shallow with a higher zero-point. Including these limits at the lower \bh\ end of the \msigma\ 
plane addresses, in part, a limitation of the sample used here. As already noted, the \sig\ catalog used here is likely incomplete due to the difficulty of measuring 
\sig\ in faint galaxies at greater distances. In addition, the \sig\ catalog likely contains inhomogenuous measurements that may not translate from bulge to 
bulge.

What are the consequences to galaxy evolution models if there is only a \msigmau\ relation? First, galaxies will no longer be required to obey the \msigma\ relation, 
and could host a SMBH with any \bh\ below \msigmau. Models that include feedback from the SMBH to the galaxy will then need to be carefully reconsidered. While the 
SMBH will undoubtedly have some influence on a portion of the host galaxy, it would not need to affect large scale properties; evolution of the SMBH would be a result 
of host galaxy evolution. An upper limit in the \msigma\ plane would also represent the pinnacle of SMBH evolution as a function of \sig, in which galaxies evolve up 
to the \msigmau\ limit. A signature of such a scenario could be a cosmic variation in $\epsilon$ (the scatter would increase with redshift) and an {\it observed} 
\msigma\ relation that does not exceed \msigmau. If the distribution of \bh\ is random within bulges, then when compared with the local {\it observed} \msigma\ 
relation, \bh\ estimates from higher redshift could fall to the left or the right. \cite{2007ApJ...667..117T} find a population of z=0.36 Seyfert 1 galaxies that 
lie above the local {\it observed} \msigma\ relation by $\Delta\log\sigma=0.13, \Delta\log$\bh$=0.54$, but this population still lies below the \msigmau\ limit 
estimated here. Finally, if SMBHs can reside anywhere below the \msigmau\ limit, then the local black hole mass function may have been over-estimated. This would 
relax the observation that merging is not important and that SMBH growth is dominated by accretion \citep[e.g.,][]{2004MNRAS.351..169M}. This potentially allows 
anti-hierarchical SMBH growth to no longer present problems for hierarchical galaxy formation models. 

\acknowledgments

The use of the HyperLeda database (http://leda.univ-lyon1.fr) is acknowledged. DB thanks Alessandro Marconi, David Merritt and Andy Robinson for useful discussions, 
and acknowledges the referee for suggesting improvements.  Support for this work was provided by proposal number HST-AR-10935.01 awarded by NASA through a grant from 
the Space Telescope Science Institute, which is operated by the Association of Universities for Research in Astronomy, Incorporated, under NASA contract NAS5-26555.


\begin{thebibliography}{0}
\expandafter\ifx\csname natexlab\endcsname\relax\def\natexlab#1{#1}\fi
\expandafter\ifx\csname bibnamefont\endcsname\relax
  \def\bibnamefont#1{#1}\fi
\expandafter\ifx\csname bibfnamefont\endcsname\relax
  \def\bibfnamefont#1{#1}\fi
\expandafter\ifx\csname citenamefont\endcsname\relax
  \def\citenamefont#1{#1}\fi
\expandafter\ifx\csname url\endcsname\relax
  \def\url#1{\texttt{#1}}\fi
\expandafter\ifx\csname urlprefix\endcsname\relax\def\urlprefix{URL }\fi
\providecommand{\bibinfo}[2]{#2}
\providecommand{\eprint}[2][]{\url{#2}}

\end{thebibliography}


\begin{thebibliography}{}
\bibitem[Adams et al.(2003)]{2003ApJ...591..125A} Adams, F.~C., Graff, D.~S., Mbonye, M., \& Richstone, D.~O.\ 2003, \apj, 591, 125 
\bibitem[Baes et al.(2003)]{2003MNRAS.341L..44B} Baes, M., Buyle, P., Hau, G.~K.~T., \& Dejonghe, H.\ 2003, \mnras, 341, L44 
\bibitem[Batcheldor \& Koekemoer(2009)]{2009PASP..121.1245B} Batcheldor, D., \& Koekemoer, A.~M.\ 2009, \pasp, 121, 1245 
\bibitem[Beifiori et al.(2009)]{2009ApJ...692..856B} Beifiori, A., Sarzi, M., Corsini, E.~M., Dalla Bont{\`a}, E., Pizzella, A., Coccato, L., \& Bertola, F.\ 2009, \apj, 692, 856 
\bibitem[Cattaneo et al.(2005)]{2005MNRAS.364..407C} Cattaneo, A., Blaizot, J., Devriendt, J., \& Guiderdoni, B.\ 2005, \mnras, 364, 407 
\bibitem[Ciotti(2008)]{2008arXiv0808.1349C} Ciotti, L.\ 2008, arXiv:0808.1349 
\bibitem[Ciotti \& van Albada(2001)]{2001ApJ...552L..13C} Ciotti, L., \& van Albada, T.~S.\ 2001, \apjl, 552, L13 
\bibitem[Coccato et al.(2006)]{2006MNRAS.366.1050C} Coccato, L., Sarzi, M., Pizzella, A., Corsini, E.~M., Dalla Bont{\`a}, E., \& Bertola, F.\ 2006, \mnras, 366, 1050 
\bibitem[Ferrarese(2002)]{2002ApJ...578...90F} Ferrarese, L.\ 2002, \apj, 578, 90 
\bibitem[Ferrarese \& Ford(2005)]{2005SSRv..116..523F} Ferrarese, L., \& Ford, H.\ 2005, Space Science Reviews, 116, 523 
\bibitem[Ferrarese \& Merritt(2000)]{2000ApJ...539L...9F} Ferrarese, L., \& Merritt, D.\ 2000, \apjl, 539, L9 
\bibitem[Gebhardt et al.(2000)]{2000ApJ...539L..13G} Gebhardt, K., et al.\ 2000, \apjl, 539, L13 
\bibitem[Graham(2008a)]{2008ApJ...680..143G} Graham, A.~W.\ 2008a, \apj, 680, 143 
\bibitem[Graham(2008b)]{2008PASA...25..167G} Graham, A.~W.\ 2008b, Publications of the Astronomical Society of Australia, 25, 167 
\bibitem[Graham et al.(2001)]{2001ApJ...563L..11G} Graham, A.~W., Erwin, P., Caon, N., \& Trujillo, I.\ 2001, \apjl, 563, L11 
\bibitem[Graham \& Li(2009)]{2009ApJ...698..812G} Graham, A.~W., \& Li, I.\ 2009, \apj, 698, 812 
\bibitem[G{\"u}ltekin et al.(2009)]{2009ApJ...698..198G} G{\"u}ltekin, K., et al.\ 2009, \apj, 698, 198 
\bibitem[H{\"a}ring \& Rix(2004)]{2004ApJ...604L..89H} H{\"a}ring, N., \& Rix, H.-W.\ 2004, \apjl, 604, L89 
\bibitem[Heckman et al.(2004)]{2004ApJ...613..109H} Heckman, T.~M., Kauffmann, G., Brinchmann, J., Charlot, S., Tremonti, C., \& White, S.~D.~M.\ 2004, \apj, 613, 109 
\bibitem[Hu(2008)]{2008MNRAS.386.2242H} Hu, J.\ 2008, \mnras, 386, 2242 
\bibitem[Kormendy \& Richstone(1995)]{1995ARA&A..33..581K} Kormendy, J., \& Richstone, D.\ 1995, \araa, 33, 581 
\bibitem[Magorrian et al.(1998)]{1998AJ....115.2285M} Magorrian, J., et al.\ 1998, \aj, 115, 2285 
\bibitem[Marconi et al.(2003)]{2003ApJ...586..868M} Marconi, A., et al.\ 2003, \apj, 586, 868 
\bibitem[Marconi \& Hunt(2003)]{2003ApJ...589L..21M} Marconi, A., \& Hunt, L.~K.\ 2003, \apjl, 589, L21 
\bibitem[Marconi et al.(2004)]{2004MNRAS.351..169M} Marconi, A., Risaliti, G., Gilli, R., Hunt, L.~K., Maiolino, R., \& Salvati, M.\ 2004, \mnras, 351, 169 
\bibitem[Mathur \& Grupe(2005)]{2005ApJ...633..688M} Mathur, S., \& Grupe, D.\ 2005, \apj, 633, 688 
\bibitem[Novak et al.(2006)]{2006ApJ...637...96N} Novak, G.~S., Faber, S.~M., \& Dekel, A.\ 2006, \apj, 637, 96 
\bibitem[Paturel et al.(2003)]{2003A&A...412...45P} Paturel, G., Petit, C., Prugniel, P., Theureau, G., Rousseau, J., Brouty, M., Dubois, P., \& Cambr{\'e}sy, L.\ 2003, \aap, 412, 45 
\bibitem[Peebles(1972)]{1972ApJ...178..371P} Peebles, P.~J.~E.\ 1972, \apj, 178, 371 
\bibitem[Pizzella et al.(2005)]{2005ApJ...631..785P} Pizzella, A., Corsini, E.~M., Dalla Bont{\`a}, E., Sarzi, M., Coccato, L., \& Bertola, F.\ 2005, \apj, 631, 785 
\bibitem[Robertson et al.(2006)]{2006ApJ...641...90R} Robertson, B., Hernquist, L., Cox, T.~J., Di Matteo, T., Hopkins, P.~F., Martini, P., \& Springel, V.\ 2006, \apj, 641, 90 
\bibitem[Sarzi et al.(2002)]{2002ApJ...567..237S} Sarzi, M., et al.\ 2002, \apj, 567, 237 
\bibitem[Treu et al.(2007)]{2007ApJ...667..117T} Treu, T., Woo, J.-H., Malkan, M.~A., \& Blandford, R.~D.\ 2007, \apj, 667, 117 
\bibitem[Valluri et al.(2004)]{2004ApJ...602...66V} Valluri, M., Merritt, D., \& Emsellem, E.\ 2004, \apj, 602, 66 
\bibitem[Vittorini et al.(2005)]{2005MNRAS.363.1376V} Vittorini, V., Shankar, F., \& Cavaliere, A.\ 2005, \mnras, 363, 1376 
\bibitem[Volonteri(2007)]{2007ApJ...663L...5V} Volonteri, M.\ 2007, \apjl, 663, L5 
\bibitem[Wyithe \& Loeb(2005)]{2005ApJ...634..910W} Wyithe, J.~S.~B., \& Loeb, A.\ 2005, \apj, 634, 910 
\end{thebibliography}
\end{document}